\documentclass[12pt,draftclsnofoot, peerreview, onecolumn]{IEEEtran}

\usepackage{gensymb}% \degree
\usepackage{balance}
\usepackage{color}
\usepackage{url}
\usepackage{bm}
\usepackage{cite}
\usepackage{algorithm}
\usepackage{threeparttable}
\usepackage{ifpdf}
\usepackage[dvips]{graphicx}
\graphicspath{{../}}
\DeclareGraphicsExtensions{.pdf,.jpeg,.png,.eps}
\ifCLASSOPTIONcompsoc
  \usepackage[tight,normalsize,sf,SF]{subfigure}
\else
  \usepackage[tight,footnotesize]{subfigure}
\fi
\usepackage{amsfonts}
\usepackage{amsmath}
\usepackage{mathrsfs}
\usepackage{multirow}
\usepackage{booktabs}
\usepackage{setspace}
\usepackage{amssymb}
\usepackage{mdwmath}
\usepackage{color}

\ifCLASSINFOpdf
\else
\fi

\newtheorem{example}{\textbf{Example}}

\ifCLASSOPTIONonecolumn
\newcommand{\figwidth}{0.65\textwidth}
\fi

\ifCLASSOPTIONtwocolumn
\newcommand{\figwidth}{0.42\textwidth}
\fi

\begin{document}
\title{Successive Cancellation List Decoding of Semi-random Unit Memory Convolutional Codes}

\author{Wenchao~Lin,
        Suihua~Cai,
        Baodian~Wei,
        and~Xiao~Ma,~\IEEEmembership{Member,~IEEE}% <-this % stops a space
\thanks{$^*$Corresponding author is Xiao Ma. This work was supported by the NSF of China~(No. 61771499 and No. 61972431), the Science and Technology Planning Project of Guangdong Province~(2018B010114001), the National Key R\&D Program of China~(2017YFB0802503), the Basic Research Project of Guangdong Provincial NSF~(2016A030308008) and the Guangdong Basic and Applied Basic Research Foundation~(2020A1515010687).}
\thanks{This work was presented in part at 2019 IEEE International Symposium on Information Theory and 2018 IEEE International Symposium on Turbo Codes \& Iterative Information Processing.}
\thanks{The authors are with the School of Data and Computer Science and Guangdong Key Laboratory of Information Security Technology, Sun Yat-sen University, Guangzhou 510006, China (e-mail: linwch7@mail2.sysu.edu.cn, caish23@mail.sysu.edu.cn, maxiao@mail.sysu.edu.cn, weibd@mail.sysu.edu.cn).}
}


\maketitle

\begin{abstract}
  We present in this paper a special class of unit memory convolutional codes~(UMCCs), called semi-random UMCCs~(SRUMCCs), where the information block is first encoded by a short block code and then transmitted in a block Markov~(random) superposition manner.
  We propose a successive cancellation list decoding algorithm, by which a list of candidate codewords are generated serially until one passes an empirical divergence test instead of the conventional cyclic redundancy check~(CRC).
  The threshold for testing the correctness of candidate codewords can be learned off-line based on the statistical behavior of the introduced empirical divergence function~(EDF).
  The performance-complexity tradeoff and the performance-delay tradeoff can be achieved by adjusting the statistical threshold and the decoding window size.
  To analyze the performance, a closed-form upper bound and a simulated lower bound are derived.
  Simulation results verify our analysis and show that:
  1) The proposed list decoding algorithm with empirical divergence test outperforms the sequential decoding in high signal-to-noise ratio~(SNR) region;
  2) Taking the tail-biting convolutional codes~(TBCC) as the basic codes, the proposed list decoding of SRUMCCs have comparable performance with the polar codes under the constraint of equivalent decoding delay.
  %In this paper, we propose a list decoding algorithm, which integrates an empirical divergence test instead of the conventional cyclic redundancy check~(CRC), for semi-random unit memory convolutional codes~(SRUMCCs).
  %The basic idea is that, any few erroneous on the first sub-frame will be boosted to the second sub-frame by the random transformation.
%  %The basic idea is that, compared with an erroneous candidate codeword, the correct candidate codeword for the first sub-frame has less effect on the decoding output of the second sub-frame.
%  The threshold for testing the correctness of the candidate codeword is then determined based on the statistical behavior of the introduced empirical divergence function~(EDF).
%  The performance-complexity tradeoff and the performance-delay tradeoff can be achieved by adjusting the statistical threshold and the decoding window size.
%  To analyze the performance, a closed-form upper bound and a simulated lower bound are derived.
%  Simulation results verify our analysis and show that:
%  1) The proposed list decoding algorithm with empirical divergence test outperforms the sequential decoding in high signal-to-noise ratio~(SNR) region;
%  2) Taking the tail-biting convolutional codes~(TBCC) as the basic codes, the proposed list decoding of SRUMCCs have comparable performance with the polar codes under the constraint of equivalent decoding delay.
\end{abstract}


%\IEEEpeerreviewmaketitle

\section{Introduction}
%\IEEEPARstart{I}{t} is well-known that reliable transmission with arbitrarily low error rate is possible with unbounded coding length as long as the transmission rate is below the channel capacity~\cite{Shannon1948Theory}.
%Although the totally random code ensemble is employed to prove the channel coding theorem, it has no efficient encoding and decoding algorithm and hence seems not suitable for practical system.
%Therefore, much effort has been expended on constructing capacity-approaching channel codes with acceptable encoding and decoding complexity.
\IEEEPARstart{T}{he} channel coding theorem states that reliable transmission with arbitrarily low error rate is possible with unbounded coding length as long as the transmission rate is below the channel capacity~\cite{Shannon1948Theory}.
The theorem was proved by the use of the random code ensemble, which usually has no efficient encoding and decoding algorithm.
Therefore, much effort has been paid on constructing capacity-approaching channel codes with acceptable encoding and decoding complexity.
Block codes and convolutional codes are two types of codes.
In block coding, the information sequence is divided into $k$-bit blocks, each being encoded independently.
A number of powerful iteratively decodable block codes with long block length have been proposed.
For example, low-density parity check~(LDPC) codes~\cite{Gallager1962ldpc} and turbo codes~\cite{Berrou1993turbo} perform near the Shannon limits under iterative belief propagation~(BP) decoding algorithm.
In contrast to block codes, the convolutional codes~\cite{Elias1955Coding} are stream-oriented.
The output from a convolutional encoder depends not only on the current input but also on the previous inputs.
The classical convolutional codes typically have small constraint length and hence perform far away from the Shannon limits.
An important class of convolutional codes is the unit memory convolutional codes~(UMCCs)~\cite{Lee1976Short}, since any convolutional code can be interpreted as a UMCC.
It was pointed out in~\cite{Lee1976Short} that the UMCCs always achieve the largest free distance among all convolutional codes with the same rate and number of encoder states, indicating that the UMCCs perform better than classical convolutional codes with the same decoding complexity.
The distance profile of the time-varying UMCCs was analyzed in~\cite{Thommesen1983Bounds} and good UMCCs were designed by search algorithms in~\cite{Ebel1996A,Said1993Using}.
Efficient decoding algorithms for UMCCs were investigated in~\cite{Hole1997Two,Freudenberger2006A,Justesen1993Bounded}.
%The authors in~\cite{Thommesen1983Bounds} analyzed the distance profile of the time-varying UMCCs and proved an asymptotic lower bound.
%In~\cite{Ebel1996A}, the author searched good $(n,k)$ UMCCs by searching the good $(2n,k)$ corresponding block codes.
%For more results on UMCCs, see~\cite{Hole1997Two,Freudenberger2006A,Said1993Using,Justesen1993Bounded}.
Since the rediscovery of LDPC codes, a class of capacity-approaching convolutional codes, called LDPC convolutional codes~\cite{Felstrom1999Time} or spatially coupled LDPC codes~\cite{Kudekar2011Threshold}, have been constructed by coupling the parity-check matrices of the LDPC block codes.
Note that the capacity-approaching convolutional codes can also be constructed by coupling the generator matrices of block codes~\cite{Ma2015Block}.
%The block Markov superposition transmission~(BMST) codes in are constructed by coupling the generator matrices of block codes.

The aforementioned codes designed for approaching the channel capacity are not suitable for emerging applications that are sensitive to the delay.
Particularly, the ultra-reliable and low latency communications~(URLLC) has caught more and more attention, which focuses on services with strict latency constraint, such as automated driving, medical applications, industrial automation and augmented/virtual reality.
Hence, it becomes important to design efficient channel codes with short and moderate length~(e.g., a thousand or less information bits)~\cite{Wonterghem2016EBCH}.
%Recently, people have paid more and more attention on the ultra-reliable and low latency communications~(URLLC), which focuses on services with strict latency constraint, such as automated driving, medical applications, industrial automation and augmented/virtual reality.
%However, the aforemention codes designed for approaching the channel capacity are not suitable for emerging applications that are sensitive to the delay.
%Designing a good code with strict latency constraint is challenging since most constructions developed for long block length do not deliver good codes in the short block length regime.
%For this reason, it becomes important to design efficient channel codes with short and moderate length~(e.g., a thousand or less information bits)~\cite{Wonterghem2016EBCH}.
One solution is to construct LDPC codes by progressive edge growth~(PEG) algorithm~\cite{Hu2005PEG}, which can deliver better codes than randomly constructed LDPC codes in short block length regime.
Polar codes~\cite{Arikan2009Polar}, another promising solution for short packet transmission, have been adopted by the 5G standard~\cite{Polar2016Std} for the control channel.
Many works on constructions, decoding algorithms and decoder implementations for short polar codes have been reported~\cite{Bai2017Polar,Tal2011SCL,Hashemi2018RMP,Yuan2014EarlyStopPolar,Zhang2013SeqPolar}.
Powerful classical short codes with near maximum likelihood decoding algorithm were also investigated for low latency communication.
In~\cite{Wonterghem2016EBCH}, the extended Bose-Chaudhuri-Hocquenghem~(BCH) codes were shown to perform near the normal approximation benchmark under ordered statistics decoding~(OSD)~\cite{Fossorier1995OSD}.
As shown in~\cite{Gaudio2017TBCC}, in the short block length regime, the tail-biting convolutional codes~(TBCCs) with the wrap-around Viterbi algorithm~(WAVA)~\cite{Shao2003WAVA} outperform significantly state-of-the-art iterative coding schemes.
For the streaming services with strict latency constraint, such as real-time online games and video conferences, convolutional codes with a small decoding window size can be alternative choices.
The comparison in~\cite{Rachinger2015ComparisonCCLDPC,Maiya2012ComparisonCCLDPC} between convolutional codes and PEG-LDPC codes showed that convolutional codes outperform LDPC codes for very short delay when the bit error rate is used as a performance metric.

In~\cite{Ma2018SRBOCC}, we have proposed a class of block oriented convolutional codes, named semi-random block oriented convolutional codes~(SRBOCCs), which is reduced to semi-random UMCCs~(SRUMCCs) if the
encoding memory $m$ is set to one.
In~\cite{Ma2019List}, taking the truncated convolutional codes as the basic codes, we proposed a list decoding algorithm for SRUMCCs.
We also showed in~\cite{Ma2019TBCC} that the performance can be further improved by taking TBCCs as the basic codes.
As extension works of~\cite{Ma2018SRBOCC,Ma2019List,Ma2019TBCC}, we present in this paper more details on the SRUMCCs.

The encoding of the SRUMCCs consists of a structured coding process and a random coding process.
At each time, the input information block~(referred to as a sub-frame) is first encoded by a structured basic code and then superimposed with the random linear transformation of the previous codeword, resulting in a sub-block for transmission.
Compared to the classical UMCCs, the input to the encoder of the SRUMCC at a time unit has the same length as the dimension of the basic code, which is typically large~(e.g., $k\geqslant32$), indicating that decoding the SRUMCCs with the Viterbi algorithm is impractical.
Also because of the block-oriented feature, it makes sense to introduce the average sub-frame error rate as the performance metric, in addition to the commonly-used bit error rate~(BER) and/or frame error rate~(FER).

Another distinguished feature of the SRUMCCs is the randomness introduced by the random linear transformation, which is critical to develop a successive cancellation list decoding algorithm.
The basic idea is to find a list of candidate codewords of the first sub-frame, and then to identify the transmitted one from the list based on the statistical behavior~(in terms of the empirical divergence) of the second sub-frame.
Evidently, any few errors in the first sub-frame will be boosted by the random transformation, resulting in a detectable effect on the second sub-frame.
Hence, the correct candidate can be reliably distinguished from the erroneous ones.

%the correct candidate can be reliably distinguished from the erroneous ones since any error pattern of the first sub-frame will result in a detectable effect on the second sub-frame due to the randomness.

%The SRUMCCs are decoded by a sliding window decoding algorithm with successive cancellation, whose performance depends critically on the performance of the first sub-frame.
%To recover the first sub-frame reliably, list decoding is conducted and the transmitted codeword is identified from the list with the help of an .
With the proposed successive cancellation list decoding, the SRUMCCs have the following three attractive features.
\begin{itemize}
  \item The construction of SRUMCCs is flexible, in the sense that any codes with fast encoding algorithms and efficient list decoding algorithms can be taken as the basic codes.
    This suggests that the SRUMCCs can support a wide range of code rates by simply choosing the basic codes with the desired rate. %adapting the basic codes to the desired rate.
  \item The performance of the successive cancellation list decoding algorithm depends critically on the performance of the first sub-frame, which can be predicted analytically by an upper bound derived from the weight enumerating functions~(WEFs) of the basic codes.
      Simulation results show that, in high SNR region, the performance of the SRUMCCs are well predicted by the upper bounds.
  %Given the weight enumerating functions~(WEFs) of the basic codes, the closed-form WEFs of the SRUMCCs can be derived for analyzing the performance.
  \item The performance-complexity tradeoff and the performance-delay tradeoff can be achieved by adjusting the statistical threshold and the decoding window size.
\end{itemize}
%As a final remark, since the SRUMCCs are block oriented  with a large block length, it makes sense to introduce the average sub-frame error rate as the performance metric.
%this is also in contrast of the classical UMCCs.

This paper is organized as follows.
In Section~\ref{SEC_2}, we present the encoding algorithm of the SRUMCCs.
In Section~\ref{SEC_3}, the list decoding with empirical divergence test is proposed.
In Section~\ref{SEC_4}, by analyzing the performance and decoding complexity, the performance-complexity tradeoff and performance-delay tradeoff are discussed.
Simulation results are presented in Section~\ref{SEC_5}.
Finally, some concluding remarks are given in Section~\ref{SEC_6}.

\section{Semi-Random Unit Memory Convolutional Code}\label{SEC_2}
\subsection{Encoding Algorithm}
\begin{figure}
  \centering
  \includegraphics[width=\figwidth]{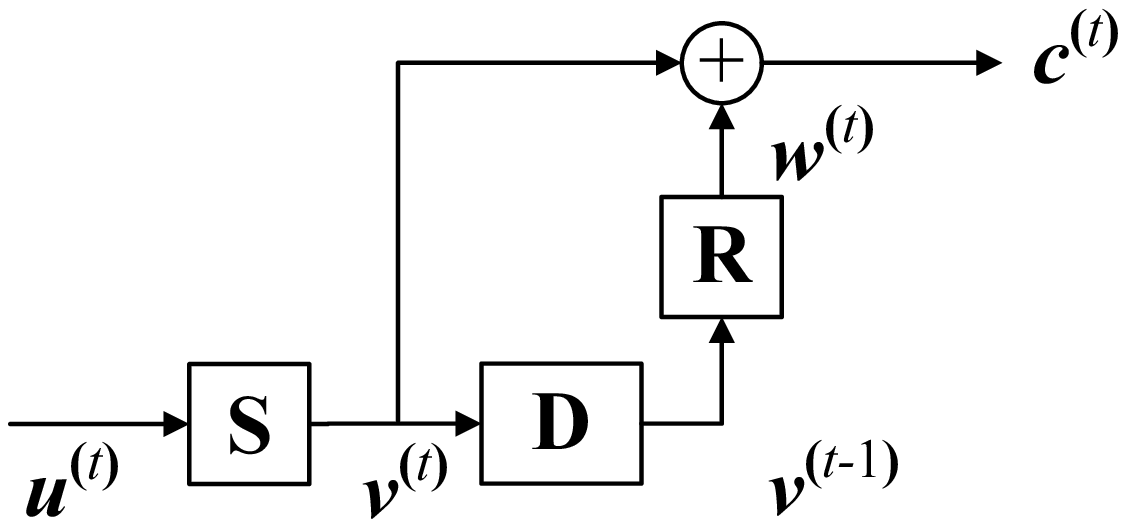}\\
  \caption{Encoding structure of the SRUMCC.}\label{FIG_ENC}
\end{figure}

Let $\boldsymbol{u} = (\boldsymbol{u}^{(0)}, \boldsymbol{u}^{(1)},\cdots,\boldsymbol{u}^{(L-1)})$ be the data to be transmitted, where $\boldsymbol{u}^{(t)}=(u^{(t)}_0,u^{(t)}_1,\cdots,u^{(t)}_{k-1})\in \mathbb{F}_2^k$ for $0\leqslant t \leqslant L-1$.
Taking a binary linear code $\mathscr{C}$ of dimension $k$ and length $n$ as the basic code, the encoding algorithm of the SRUMCC is described in Algorithm~\ref{EncodingAlgorithm}~(see Fig.~\ref{FIG_ENC} for reference).
The code rate of the SRUMCC is $R = k/n\times L/(L+1)$, which is slightly less than that of the basic code $\mathscr{C}$.
However, the rate loss is negligible for large $L$.

\begin{algorithm}\caption{Encoding of the SRUMCC}\label{EncodingAlgorithm}
\begin{itemize}
  \item{\textbf{Initialization:}} \label{step:encoding_initialize} Let $\boldsymbol{v}^{(-1)} = \boldsymbol{0} \in \mathbb{F}_2^{n}$.
  \item{\textbf{Iteration:}} \label{step:encoding_loop} For $0\leqslant t \leqslant L-1$,
      \begin{enumerate}
        \item \emph{Structured Encoding}: Encode $\boldsymbol{u}^{(t)}$ into $\boldsymbol{v}^{(t)}\in \mathbb{F}_2^n$ by the encoding algorithm of the basic code $\mathscr{C}$. Equivalently, $\boldsymbol{v}^{(t)} = \boldsymbol{u}^{(t)}\mathbf{S}$, where $\mathbf{S}$ is the generator matrix of the basic code $\mathscr{C}$.
        \item \emph{Random Transformation}: Compute $\boldsymbol{w}^{(t)} = \boldsymbol{v}^{(t-1)}\mathbf{R}\in \mathbb{F}_2^n$, where $\mathbf{R}$ is a random generated but fixed matrix of order $n$ whose elements are generated independently according to the Bernoulli distribution with success probability $1/2$.
        \item \emph{Superposition}: Compute $\boldsymbol{c}^{(t)} = \boldsymbol{v}^{(t)}+\boldsymbol{w}^{(t)}\in \mathbb{F}_2^{n}$, which will be taken as the sub-frame for transmission at time $t$.
      \end{enumerate}
  \item{\textbf{Termination:}} \label{step:encoding_termination}
        The sub-frame at time $L$ is set to $\boldsymbol{c}^{(L)} = \boldsymbol{v}^{(L-1)}\mathbf{R}$, which is equivalent to setting $\boldsymbol{u}^{(L)} = \boldsymbol{0}$.
\end{itemize}
\end{algorithm}

\textbf{Remarks:}  Recalling that the encoding of classical UMCCs is performed by %initializ $\boldsymbol{u}^{(-1)}=\boldsymbol{0}$ and to
computing $\boldsymbol{c}^{(t)}=\boldsymbol{u}^{(t)}\mathbf{G}_0+\boldsymbol{u}^{(t-1)}\mathbf{G}_1$ for $t \geqslant 0$, the proposed SRUMCCs
%The proposed SRUMCCs, whose generator matrices~(terminated) can be written as
%\begin{equation}
%\mathbf{G} = \left(
%     \begin{array}{cccccccc}
%       \mathbf{S}  & \mathbf{S}\mathbf{R} &                      &                  &\\
%                   & \mathbf{S}           & \mathbf{S}\mathbf{R} &                  &\\
%                   &                      &     \ddots           &\ddots            &\\
%                   &                      &                      & \mathbf{S}       &\mathbf{S}\mathbf{R}
%     \end{array}
%   \right),
%\end{equation}
can be viewed as a special class of UMCCs with $\mathbf{G}_0=\mathbf{S}$ and $\mathbf{G}_1=\mathbf{SR}$, of which one is structured and the other is random, hence the name.
The speciality is outlined as below.
\begin{itemize}
  \item Unlike commonly accepted classical UMCCs with small $k$, the SRUMCCs typically have large $k$~(hence large constraint length) induced by the block oriented encoding process, as is the same case for the convolutional LDPC codes. It makes sense to introduce the average sub-frame error rate, which is denoted as fER and described in the next subsection, as a new performance metric.
  \item Due to the large constraint length, the Viterbi algorithm~(VA), which is an efficient maximum likelihood decoding algorithm for classical UMCCs, does not apply to the decoding of SRUMCCs. Therefore, it is important to develop an efficient decoding algorithm for SRUMCCs, which is the main topic of this paper.
  \item The encoding of the SRUMCCs involves a structured coding process and a random coding process, thus termed as ``semi-random''. Good UMCCs with short constraint length are usually constructed by computer search~\cite{Ebel1996A}, while good SRUMCCs can be constructed easily by generating $\mathbf{R}$ randomly. The randomness is helpful to the decoding process of the SRUMCCs, since any error pattern of the first sub-frame will result in a detectable effect on the next sub-frame.
\end{itemize}

%Unlike commonly accepted classical UMCCs with small $k$, the SRUMCCs typically have large $k$~(hence large constraint length) induced by the block oriented encoding process, as is the same case for the convolutional LDPC codes. Due to the large constraint length, the SRUMCCs are typically non-decodable by the Viterbi algorithm~(VA). Another difference between the SRUMCCs and classical convolutional codes is the randomness introduced by the random linear transformation. Good convolutional codes with short constraint length are usually constructed by computer search~\cite{Lin1983Error}, while good SRUMCCs can be constructed easily by generating $\mathbf{R}$ randomly.

As a kind of convolutional code, the SRUMCC has streaming properties. In other words, the encoded bits can be generated without waiting for the whole input block while the received signal can be decoded by a sliding window decoding algorithm with tunable delays. In contrast to block codes, in the SRUMCC coded system, the latency constraint is fulfilled by the limited decoding window instead of the short coding length.

\subsection{Performance Metric}\label{SUBSEC_ENCPERMET}
Suppose that $\boldsymbol{c}^{(t)}$ is modulated with binary phase-shift keying~(BPSK) signals and transmitted over additive white Gaussian noise~(AWGN) channels, resulting in a noisy version $\boldsymbol{y}^{(t)}\in \mathbb{R}^n$ at the receiver. We focus on a sliding window decoding algorithm with the decoding window $w$, which attempts to recover $\boldsymbol{u}^{(t)}$ from $(\boldsymbol{y}^{(t)},\cdots,\boldsymbol{y}^{(t+w-1)})$. In other words, the decoding delay is $wn$ in terms of bits.

Given a decoding algorithm, define ${\rm fER}_t$ for $0\leqslant t\leqslant L-1$ as the probability that the decoding result $\hat{\boldsymbol{u}}^{(t)}$ is not equal to the transmitted vector $\boldsymbol{u}^{(t)}$ and $\rm FER$ as the probability that the decoding result $\hat{\boldsymbol{u}}$ is not equal to $\boldsymbol{u}$. It is not difficult to verify that
\begin{equation}
{\rm fER}_0\leqslant \max_{t}{\rm fER}_t \leqslant {\rm FER} \leqslant \sum_{t=0}^{L-1}{\rm fER}_t.
\end{equation}
We define
\begin{equation}\label{fER}
{\rm fER} = \frac{1}{L}\sum_{t=0}^{L-1}{\rm fER}_t,
\end{equation}
which is used as the performance metric in this paper\footnote{For conventional block codes, such as polar codes, we define $\rm fER$ as the probability that the decoding codeword is not equal to the transmitted codeword. That is, $\rm fER = FER$.} and can be evaluated in practice by
\begin{equation}\label{fER2}
{\rm fER} = \frac{\textrm{number of erroneous decoded sub-frames}} {\textrm{total number of transmitted sub-frames}}.
\end{equation}
%Given a decoding algorithm, define subFER as the probability that the decoding result $\hat{\boldsymbol{u}}^{(0)}$ is not equal to the transmitted vector $\boldsymbol{u}^{(0)}$ and FER as the probability that the decoding result $\hat{\boldsymbol{u}}$ is not equal to $\boldsymbol{u}$. Clearly, we have
%\begin{equation}\label{bound}
%  {\rm subFER} \leqslant {\rm FER} \leqslant L\cdot {\rm subFER}.
%\end{equation}
%In practice, we define
%\begin{equation}\label{fER}
%  {\rm fER} = \frac{\textrm{number of erroneous decoded sub-frames}} {\textrm{total number of transmitted sub-frames}},
%\end{equation}
%which is used as the performance metric in this paper\footnote{For block code with short block length, we define fER as the probability that the decoding codeword is not equal to the transmitted codeword. That is, fER = FER.}.

The event that the decoding result $\hat{\boldsymbol{u}}^{(0)}$ is not equal to the transmitted vector $\boldsymbol{u}^{(0)}$ is referred to as the {\em first error event} $E_0$. In general, we say that the first error event occurs at time $t$, which is denoted by $E_t$, if $\hat{\boldsymbol{u}}^{(i)} = \boldsymbol{u}^{(i)}$ for all $i < t$ but $\hat{\boldsymbol{u}}^{(t)} \neq \boldsymbol{u}^{(t)}$. The probability that the first error event occurs at time $t$ can be bounded by
\begin{equation}
{\rm Pr}\{E_t\} \leqslant {\rm fER}_0,
\end{equation}
since with $\hat{\boldsymbol{u}}^{(t-1)}$ being correct, the performance of the $t$-th sub-frame will not be worse than that of the first sub-frame.
%The probability that the first error event occurs at time $t$ is given by
%\begin{equation}
%{\rm Pr}\{E_t\} = {\rm fER}_0\cdot(1-{\rm fER}_0)^t,
%\end{equation}
%by noticing that ${\rm Pr}\{\hat{\boldsymbol{u}}^{(i)} \neq \boldsymbol{u}^{(i)}|\hat{\boldsymbol{u}}^{(i-1)} = \boldsymbol{u}^{(i-1)}\} = {\rm fER}_0$ for $1\leqslant i\leqslant L-1$.
%
%\begin{align}
%{\rm Pr}\{E_t\} &=
%{\rm Pr}\{\hat{\boldsymbol{u}}^{(t)} \neq \boldsymbol{u}^{(t)}|\hat{\boldsymbol{u}}^{(t-1)} = \boldsymbol{u}^{(t-1)}\}
%{\rm Pr}\{\hat{\boldsymbol{u}}^{(0)} = \boldsymbol{u}^{(0)}\}
%\prod_{i=1}^{t-1}{\rm Pr}\{\hat{\boldsymbol{u}}^{(i)} = \boldsymbol{u}^{(i)}|\hat{\boldsymbol{u}}^{(i-1)} = \boldsymbol{u}^{(i-1)}\}\\
%&={\rm fER}_0\cdot(1-{\rm fER}_0)^t
%\end{align}
In the worst case, the first error event at time $t$ causes catastrophic error-propagation. That is, the event $E_t=\{\hat{\boldsymbol{u}}^{(t)} \neq \boldsymbol{u}^{(t)}\}$ can cause $\hat{\boldsymbol{u}}^{(j)} \neq \boldsymbol{u}^{(j)}$ for all $j > t$. The ${\rm fER}_t$ can be bounded by
\begin{align}
{\rm fER}_t &= \sum_{i=0}^{t}{\rm Pr}\{E_i\}{\rm Pr}\{\hat{\boldsymbol{u}}^{(t)} \neq \boldsymbol{u}^{(t)}|E_i\} \nonumber \\
&\leqslant \sum_{i=0}^{t}{\rm Pr}\{E_i\}=(t+1){\rm fER}_0.
\end{align}

Therefore, the $\rm fER$ can be upper bounded by
%\begin{align}
%{\rm fER} & \leqslant \frac{1}{L}\sum_{t=0}^{L-1}(t+1){\rm fER}_0\nonumber \\
%&=\frac{L+1}{2} \cdot {\rm fER}_0.\label{bound_fer}
%\end{align}
\begin{equation}
{\rm fER}  \leqslant \frac{1}{L}\sum_{t=0}^{L-1}(t+1){\rm fER}_0=\frac{L+1}{2} \cdot {\rm fER}_0.\label{bound_fer}
\end{equation}
\section{Successive Cancellation List Decoding}\label{SEC_3}
%The SRBO-CC can be decoded by a sliding window algorithm with successive cancellation.
As a kind of convolutional codes with large constraint length, the SRUMCCs are typically non-decodable by VA. The sub-optimal sequential decoding mentioned in~\cite{Ma2018SRBOCC} can be employed for decoding the SRUMCCs, although the memory load is heavy due to the requirement of a large amount of stack memory.

In this paper, we propose a sliding window algorithm with successive cancellation. The first and critical step is to recover reliably $\boldsymbol{v}^{(0)}$, which is not interfered by any other sub-frames. By removing the effect of the first sub-frame, the second sub-frame is then decoded in the same way. This process will be continued until all sub-frames are decoded. In this section, we focus on the methods to estimate $\boldsymbol{v}^{(0)}$ from $(\boldsymbol{y}^{(0)},\boldsymbol{y}^{(1)})$. The complete decoding algorithm is summarized in Algorithm~\ref{DecodingAlgorithmW2} and the extension to the recovery of $\boldsymbol{v}^{(0)}$ from $(\boldsymbol{y}^{(0)},\boldsymbol{y}^{(1)},\boldsymbol{y}^{(2)})$ will be discussed in Subsection~\ref{SUBSEC_DECW3}.

For illustrating the basic idea, we first introduce the maximum a posteriori~(MAP) decoding and the maximum likelihood~(ML) decoding, although they seem to be less practical. The list decoding with empirical divergence test is then proposed.

\subsection{Maximum A Posteriori Decoding}
The MAP decoding is optimal in the sense that the error probability of $\boldsymbol{v}^{(0)}$~(i.e., $\rm fER_0$) is minimized. The MAP decoder always outputs the codeword\footnote{Without causing much ambiguity, we will use $\hat{\boldsymbol{v}}^{(0)}$, $\hat{\boldsymbol{c}}^{(0)}$ and $\hat{\boldsymbol{u}}^{(0)}$ interchangeably in the remainder of this paper.}
\begin{align}
\hat{\boldsymbol{v}}^{(0)} = &\mathop{\arg\max_{\boldsymbol{v}^{(0)}}}P(\boldsymbol{v}^{(0)}|\boldsymbol{y}^{(0)}\boldsymbol{y}^{(1)}) \nonumber \\
= &\mathop{\arg\max_{\boldsymbol{v}^{(0)}}}
\frac{P(\boldsymbol{v}^{(0)})}{P(\boldsymbol{y}^{(0)}\boldsymbol{y}^{(1)})}P(\boldsymbol{y}^{(0)}\boldsymbol{y}^{(1)}|\boldsymbol{v}^{(0)}),
\end{align}
where $P(\cdot)$ is the probability mass (or density) function.
Since the channel is memoryless and $\boldsymbol{v}^{(0)}$ is independent with $\boldsymbol{v}^{(1)}$, we have
\begin{align}
P(\boldsymbol{y}^{(0)}\boldsymbol{y}^{(1)}|\boldsymbol{v}^{(0)})
= &P(\boldsymbol{y}^{(0)}|\boldsymbol{v}^{(0)})P(\boldsymbol{y}^{(1)}|\boldsymbol{v}^{(0)})\nonumber\\
= &P(\boldsymbol{y}^{(0)}|\boldsymbol{v}^{(0)})
\sum_{\boldsymbol{v}^{(1)}}P(\boldsymbol{v}^{(1)})P(\boldsymbol{y}^{(1)}|\boldsymbol{v}^{(0)}\boldsymbol{v}^{(1)}).
\end{align}
Therefore, we have
\begin{equation}
P(\boldsymbol{v}^{(0)}|\boldsymbol{y}^{(0)}\boldsymbol{y}^{(1)})\varpropto P(\boldsymbol{y}^{(0)}|\boldsymbol{v}^{(0)})\sum_{\boldsymbol{v}^{(1)}}P(\boldsymbol{y}^{(1)}|\boldsymbol{v}^{(0)}\boldsymbol{v}^{(1)}),
\end{equation}
by noticing that $\frac{P(\boldsymbol{v}^{(0)})}{P(\boldsymbol{y}^{(0)}\boldsymbol{y}^{(1)})}$ is constant for all $\boldsymbol{v}^{(0)}$.

To find such a codeword $\hat{\boldsymbol{v}}^{(0)}$, the MAP decoder explores all $2^{2nR}$ possible codewords $(\boldsymbol{v}^{(0)},\boldsymbol{v}^{(1)})$, which implies that the complexity increases exponentially with the length of the basic code.

\subsection{Maximum Likelihood Decoding}
Different from the MAP decoder, the ML decoder minimizes the error probability of the codeword $(\boldsymbol{v}^{(0)},\boldsymbol{v}^{(1)})$. The ML decoder selects $\hat{\boldsymbol{v}}^{(0)}$ as output such that
\begin{equation}
(\hat{\boldsymbol{v}}^{(0)},\hat{\boldsymbol{v}}^{(1)}) = \mathop{\arg\max_{(\boldsymbol{v}^{(0)},\boldsymbol{v}^{(1)})}}
P(\boldsymbol{y}^{(0)}\boldsymbol{y}^{(1)}|\boldsymbol{v}^{(0)}\boldsymbol{v}^{(1)}).
\end{equation}
Since the channel is memoryless, we have
\begin{equation}
P(\boldsymbol{y}^{(0)}\boldsymbol{y}^{(1)}|\boldsymbol{v}^{(0)}\boldsymbol{v}^{(1)})=
P(\boldsymbol{y}^{(0)}|\boldsymbol{v}^{(0)})P(\boldsymbol{y}^{(1)}|\boldsymbol{v}^{(0)}\boldsymbol{v}^{(1)}).
\end{equation}
Equivalently, the ML decoder outputs the codeword
\begin{equation}
\hat{\boldsymbol{v}}^{(0)} = \mathop{\arg\max_{\boldsymbol{v}^{(0)}}}
P(\boldsymbol{y}^{(0)}|\boldsymbol{v}^{(0)})\left[\max_{\boldsymbol{v}^{(1)}}P(\boldsymbol{y}^{(1)}|\boldsymbol{v}^{(0)}\boldsymbol{v}^{(1)})\right].\label{ML_METRIC}
\end{equation}
The ML decoding can also be viewed as an approximation to the MAP decoding, since the term
 $\max_{\boldsymbol{v}^{(1)}}P(\boldsymbol{y}^{(1)}|\boldsymbol{v}^{(0)}\boldsymbol{v}^{(1)})$ is the dominant term in $\sum_{\boldsymbol{v}^{(1)}}P(\boldsymbol{y}^{(1)}|\boldsymbol{v}^{(0)}\boldsymbol{v}^{(1)})$.

Given $\boldsymbol{v}^{(0)}$, the inner maximization over $\boldsymbol{v}^{(1)}$ in~(\ref{ML_METRIC}) can be achieved by performing the Viterbi algorithm~(VA), which is more efficient than exploring all possible $\boldsymbol{v}^{(1)}$. Unfortunately, no efficient algorithm to achieve the outer maximization, except exploring all $2^{nR}$ possible $\boldsymbol{v}^{(0)}$, which implies that the complexity is lower than the MAP decoding but still increases exponentially with the length of the basic code.

\subsection{List Decoding with Empirical Divergence Test}\label{SUBSEC_DECW2}
One obvious way to reduce the complexity of the ML decoding is to limit the search space for $\boldsymbol{v}^{(0)}$. Let $\mathcal{L}\subset\mathscr{C}$ be a list of $\ell_{\rm max}$ codewords. The decoder outputs the codeword
\begin{equation}
\hat{\boldsymbol{v}}^{(0)} = \mathop{\arg\max_{\boldsymbol{v}^{(0)}\in\mathcal{L}}}
P(\boldsymbol{y}^{(0)}|\boldsymbol{v}^{(0)})\left[\max_{\boldsymbol{v}^{(1)}}P(\boldsymbol{y}^{(1)}|\boldsymbol{v}^{(0)}\boldsymbol{v}^{(1)})\right].
\end{equation}
Obviously, if the transmitted codeword $\boldsymbol{v}^{(0)}$ is included in the list $\mathcal{L}$, the decoder with reduced search space performs no worse than the ML decoder. In contrast, an error must occur if the transmitted codeword is not in the list. Therefore, we need to generate efficiently a list $\mathcal{L}$ which contains the transmitted codeword with high probability.

We assume that the basic code $\mathscr{C}$ can be efficiently decoded by outputting a list of candidate codewords. To avoid messy notation, we omit the superscript of $\boldsymbol{v}^{(0)}$ and assume that a codeword $\boldsymbol{v}\in \mathscr{C}$ is transmitted. Upon receiving its noisy version $\boldsymbol{y}= (y_0, y_1, \cdots, y_{n-1})$, the decoder {\em serially} outputs a list of candidate codewords $\hat{\boldsymbol{v}}_{\ell}$, $\ell = 1, 2, \cdots, \ell_{\max}$, where $\ell_{\max}$ is a parameter to trade off the performance against the complexity.
We will not focus on the detailed implementation in this paper but simply conduct the serial list Viterbi algorithm~(SLVA)~\cite{Seshadri1994LVA} over the trellis representation of the basic code. For ease of notation, we use SLVA($\boldsymbol{y}$, $\ell$) to represent the $\ell$-th output of the SLVA. In particular, SLVA($\boldsymbol{y}$, 1), simply denoted by VA($\boldsymbol{y}$), is the output of the VA.

%For any binary vector $\boldsymbol{x}$, its likelihood is given by $f(\boldsymbol{y}|\boldsymbol{x}) = \prod_{i=0}^{n-1} f(y_i|x_i)$, where $f(y_i|x_i)$ is the considered conditional probability density function specified by the modulation and the channel. By the nature of the SLVA, we have $f(\boldsymbol{y}|\hat{\boldsymbol{v}}_1) \geqslant  f(\boldsymbol{y}|\hat{\boldsymbol{v}}_2) \geqslant \cdots \geqslant f(\boldsymbol{y}|\hat{\boldsymbol{v}}_{\ell_{\max}})$, where $\hat{\boldsymbol{v}}_\ell$=SLVA($\boldsymbol{y}$, $\ell$).
The list decoding is {\em successful} if the transmitted codeword occurs in the list. Obviously, the probability of the list decoding being successful can be as high as required by enlarging the list size $\ell_{\max}$. \textbf{Example}~\ref{EX_LIST} shows the performance of a TBCC under list decoding.

\begin{figure}
  \centering
  \includegraphics[width=\figwidth]{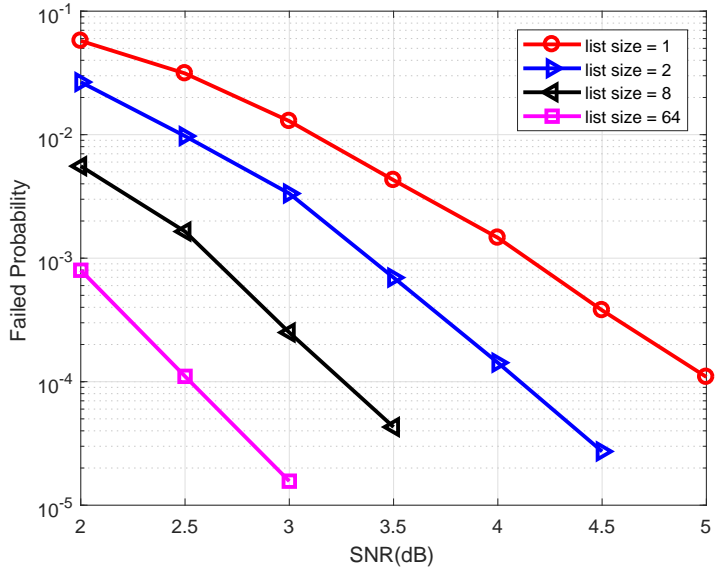}\\
  \caption{Performance of the TBCC under list decoding in \textbf{Example}~\ref{EX_LIST}. The $16$-state $(2,1,4)$ TBCC defined by the polynomial generator matrix~\cite{Lin1983Error} $G(D) = [D^4+D^2+D+1,D^4+D^3+1]$~(denoted as $[27, 31]_8$ in octal form for short) with information length $k=32$~($n=64$) is considered. Here the label ``Failed Probability'' for the ordinate stands for the probability that the transmitted codeword is not contained in the list.}\label{FIG_LIST}
\end{figure}

\begin{example}\label{EX_LIST}
The $16$-state $(2,1,4)$ TBCC defined by the polynomial generator matrix~\cite{Lin1983Error} $G(D) = [D^4+D^2+D+1,D^4+D^3+1]$~(denoted as $[27, 31]_8$ in octal form for short) with information length $k=32$~($n=64$) is considered. The list decoding performance is shown in Fig.~\ref{FIG_LIST}.
%and the average list size required to contain the correct candidate codeword is given in Table~\ref{TAB_LIST}.
%We observe that the performance can be improved by increasing the list size.
\end{example}

\begin{table}[tp]
  \centering
  \caption{Average list sizes required to contain the correct candidate codeword}
  \begin{tabular}{|l|l|l|l|l|l|l|}\hline
  ${\rm SNR}$   & 2.0 & 2.5 & 3.0 & 3.5 & 4.0 \\\hline
  list size     &1.256&1.069&1.019&1.005&1.001\\\hline
  \end{tabular}\label{TAB_LIST}
\end{table}

%Let $\ell^*$ be the number of codewords that are more likely than the transmitted one. In \textbf{Example}~\ref{EX_LIST}, we also calculate $\mathbb{E}[\ell^*]$, the average list size needed to include the transmitted codeword. It should be pointed out that, for a large $\ell_{\rm max}$, $\ell^*$ is far less than $\ell_{\rm max}$ with high probability.

%For example, from Fig.~\ref{FIG_LIST}, at $3~{\rm dB}$, the probability that $\ell^*\leqslant8\ll\ell_{\rm max}=64$ is about $(1-2\times10^{-4})$.

For a large list size~(e.g., $\ell_{\rm max}=64$), the transmitted codeword is included in the list with high probability. However, the average list sizes required to contain the correct candidate codeword can be much smaller than $\ell_{\rm max}=64$, as tabulated in Table~\ref{TAB_LIST}. This implies that, in many cases, the list size can be smaller than $\ell_{\rm max}$.
To reduce the complexity, a \emph{serial} list decoding is employed for the basic codes so that we can exit the decoding algorithm once the correct candidate codeword is identified.
Then a question arises: How to check the correctness of the candidate codeword? One solution is to invoke the cyclic redundancy check~(CRC), as embedded in polar codes~\cite{Niu2012CRCPolar}. However, the overhead~(rate loss) due to the CRC is intolerable especially for a short basic code. Motivated by the jointly typical set decoding, which is employed in~\cite[{Section} 3.2]{Han2002Information} to prove the channel coding theorem, we consider checking the correctness of the candidate codeword by typicality. The list decoding process will terminate if a candidate codeword is found to be ``jointly typical'' with the received signal.
%This motivates us to consider checking the correctness of the candidate codeword by typicality. The list decoding process will stop if a codeword which is ``jointly typical'' with the received signal is found.

To proceed, we need the following concept. For the received signal $\boldsymbol{y}=(y_0,\cdots,y_{n-1})\in\mathbb{R}^n$, we define an {\em empirical divergence function}~(EDF) as
\begin{equation}
D(\boldsymbol{x}, \boldsymbol{y}) = \frac{1}{n} \log_2 \frac{P(\boldsymbol{y}|\boldsymbol{x})}{P(\boldsymbol{y})},
\end{equation}
for $\boldsymbol{x} \in \mathbb{F}_2^n$, where
\begin{equation}
P(\boldsymbol{y}) = \prod_{i=0}^{n-1} \left(\frac{1}{2}P(y_i|0)+\frac{1}{2}P(y_i|1)\right).
\end{equation}
Note that, in the above definition, $P(\boldsymbol{y})$ is not equal to $2^{-k}\sum_{\boldsymbol{v}\in \mathscr{C}}P(\boldsymbol{y}|\boldsymbol{v})$ but to $2^{-n}\sum_{\boldsymbol{x}\in \mathbb{F}_2^n}P(\boldsymbol{y}|\boldsymbol{x})$. Also note that $\boldsymbol{x}$ is not necessarily a codeword of $\mathscr{C}$. Especially, we are interested in the following cases.

%With finite coding length and a specified basic code which may not be well designed, we investigate the statistical behavior of the  empirical divergence function and modify the criterion of the jointly typical set decoding. To show the statistical behavior of the  empirical divergence function, we take the following cases as examples.

%In the following, we focus on the statistical behavior of the divergence function in the following cases.

\begin{enumerate}
  \item If $\boldsymbol{v}$ is the transmitted codeword, we have $D(\boldsymbol{v}, \boldsymbol{y})\approx I(X; Y) > 0$, where $\approx$ is used to indicate that the EDF is around in probability its expectation for large $n$. Here $I(X; Y)$ is the mutual information between the channel output $Y$ and the uniform binary input $X$. To be precise, $D(\boldsymbol{v}, \boldsymbol{y})\approx I(X; Y)$ means that, for an arbitrary small positive number $\epsilon$,
      \begin{equation}
      \lim_{n\rightarrow\infty}P\left[\big|D(\boldsymbol{v}, \boldsymbol{y})-I(X; Y)\big|\leqslant\epsilon\right]=1,
      \end{equation}
      as guaranteed by the weak law of large numbers~(WLLN).
%      This is actually implied by the weak law of large numbers
%      \begin{equation}
%      \lim_{n\rightarrow\infty}P\left[\big|\sum_{i=1}^nW_n-\mu\big|<\epsilon\right]=1,
%      \end{equation}
%      where $W_1,\cdots,W_n$ are a sequence of independent identically distributed random variables with finite mean $E[W]=\mu$ and $\epsilon$ is arbitrary small positive number.
  \item If $\boldsymbol{x}$ is randomly generated~(hence typically not equal to the transmitted codeword), from the WLLN, we have
      \begin{equation}
      D(\boldsymbol{x}, \!\boldsymbol{y})\!\approx\! \mathbb{E}_{Y\!|\!V}\!\left[\frac{1}{2}\log_2\!\frac{P(Y|0)}{P(Y)}\!+\!\frac{1}{2}\log_2\!\frac{P(Y|1)}{P(Y)} \right],
      \end{equation}
      which is negative from the concavity of the function $\log_2(\cdot)$.
  \item What are the typical values of $D(\hat{\boldsymbol{v}}, \boldsymbol{y})$, where $\hat{\boldsymbol{v}}={\rm VA}(\boldsymbol{y})$? Given $\boldsymbol{y}$, since $D(\hat{\boldsymbol{v}}, \boldsymbol{y}) = \max_{\boldsymbol{v}\in \mathscr{C}}D(\boldsymbol{v},\boldsymbol{y})$, we expect that $D(\hat{\boldsymbol{v}}, \boldsymbol{y})\geqslant D(\boldsymbol{v}, \boldsymbol{y})\approx I(X; Y) > 0$.
  \item What about $D(\tilde{\boldsymbol{v}}, \tilde{\boldsymbol{y}})$? Here $\tilde{\boldsymbol{v}} = {\rm VA}(\tilde{\boldsymbol{y}})$ where $\tilde{\boldsymbol{y}} = \boldsymbol{x} \odot \boldsymbol{y}$ with $\boldsymbol{x}$ being a totally random bipolar vector and $\odot$ stands for component-wise product. That is, we first randomly flip the received vector $\boldsymbol{y}$, and then execute the VA to find the first candidate codeword $\tilde{\boldsymbol{v}}$. We expect that $D(\tilde{\boldsymbol{v}}, \tilde{\boldsymbol{y}})$ is located between $D(\boldsymbol{v}, \boldsymbol{y})$ of the first case and $D(\boldsymbol{x}, \boldsymbol{y})$ of the second case.
      %We expect that $D(\tilde{\boldsymbol{v}}_1, \tilde{\boldsymbol{y}})$ is smaller than $D(\boldsymbol{v}, \boldsymbol{y})$ discussed in the first case.

\end{enumerate}

\begin{figure}
  \centering
  \includegraphics[width=\figwidth]{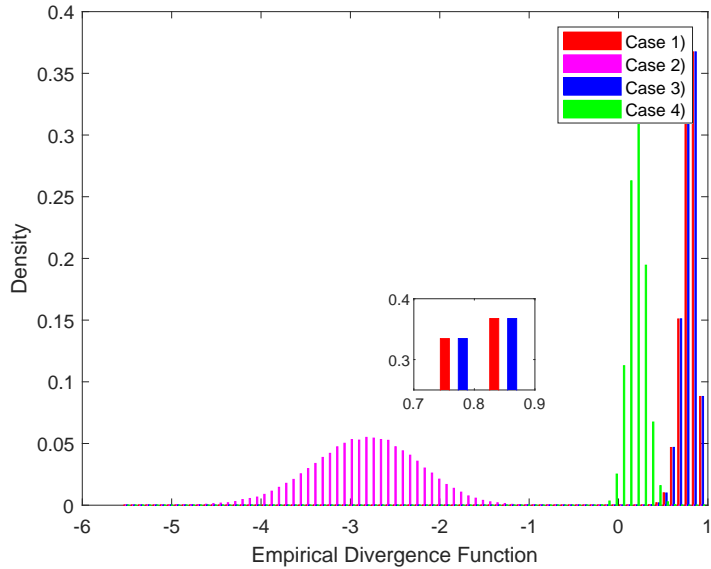}\\
  \caption{Statistical behavior of $D(\boldsymbol{x},\boldsymbol{y})$ in \textbf{Example}~\ref{EX_STATISTIC}. We consider the $16$-state $(2,1,4)$ TBCC defined by the polynomial generator matrix $G(D) = [27, 31]_8$ with information length $k=32$~($n=64$) and the AWGN channel with ${\rm SNR} = 4~{\rm dB}$, at which the mutual information is $I(X; Y)\approx0.79$.}\label{FIG_STATISTIC}
\end{figure}

\begin{example}\label{EX_STATISTIC}
We consider the TBCC in \textbf{Example}~\ref{EX_LIST} again and set ${\rm SNR} = 4~{\rm dB}$, at which the mutual information is $I(X; Y)\approx0.79$. The histogram is shown in Fig.~\ref{FIG_STATISTIC}, from which we observed that $D(\boldsymbol{v}, \boldsymbol{y})$ is likely to be large with $\boldsymbol{v}$ being the transmitted codeword~(or the output of the VA corresponding to $\boldsymbol{y}$). Note that the statistical behavior of $D(\tilde{\boldsymbol{v}}, \tilde{\boldsymbol{y}})$ is different from that of $D(\boldsymbol{x}, \boldsymbol{y})$, since $\tilde{\boldsymbol{v}}$ is dependent on $\tilde{\boldsymbol{y}}$. The typical values of $D(\tilde{\boldsymbol{v}}, \tilde{\boldsymbol{y}})$ are greater than those of $D(\boldsymbol{x}, \boldsymbol{y})$ but less than those of $D(\boldsymbol{v}, \boldsymbol{y})$.
\end{example}

The statistical behavior of the EDF can be helpful in the decoding process of the SRUMCCs. In the case when the decoding result of the first sub-frame $\hat{\boldsymbol{v}}^{(0)}$ equals to $\boldsymbol{v}^{(0)}$, $\boldsymbol{y}^{(1)}\odot\phi(\hat{\boldsymbol{v}}^{(0)}\mathbf{R})$ is the Gaussian noisy version of $\boldsymbol{v}^{(1)}$, where $\phi(\hat{\boldsymbol{v}}^{(0)}\mathbf{R})$ is the BPSK signal corresponding to the binary vector $\hat{\boldsymbol{v}}^{(0)}\mathbf{R}$. In contrast, in the case when $\hat{\boldsymbol{v}}^{(0)}\neq\boldsymbol{v}^{(0)}$, $\boldsymbol{y}^{(1)}\odot\phi(\hat{\boldsymbol{v}}^{(0)}\mathbf{R})$ is the randomly flipped Gaussian noisy version of $\boldsymbol{v}^{(1)}$. Since these two cases have different statistical impact on the EDF, we are able to distinguish with high probability whether $\boldsymbol{y}^{(1)}\odot\phi(\hat{\boldsymbol{v}}^{(0)}\mathbf{R})$ is the randomly flipped Gaussian noisy version of $\boldsymbol{v}^{(1)}$~(equivalently, $\hat{\boldsymbol{v}}^{(0)}$ is erroneous) or not.

Given $\boldsymbol{y}^{(0)}$, the SLVA is implemented to deliver serially a list of candidate codewords $\hat{\boldsymbol{v}}^{(0)}_\ell$,  for $1\leqslant \ell \leqslant \ell_{\rm max}$. For each candidate codeword, we define a soft metric
\begin{equation}\label{SOFT_METRIC}
M_{2}(\hat{\boldsymbol{v}}^{(0)}_\ell) = D(\hat{\boldsymbol{v}}^{(0)}_\ell,\boldsymbol{y}^{(0)}) + D(\tilde{\boldsymbol{v}}_\ell,\boldsymbol{y}^{(1)}\odot\phi(\hat{\boldsymbol{v}}^{(0)}_\ell\mathbf{R})),
\end{equation}
where $\tilde{\boldsymbol{v}}_\ell$ is the output of the VA with $\boldsymbol{y}^{(1)}\odot\phi(\hat{\boldsymbol{v}}^{(0)}_\ell\mathbf{R})$ as the input. The first term in the right hand side of~(\ref{SOFT_METRIC}) specifies the EDF between the candidate codeword and the received vector $\boldsymbol{y}^{(0)}$, while the second term is the EDF between $\boldsymbol{y}^{(1)}\odot\phi(\hat{\boldsymbol{v}}^{(0)}_\ell\mathbf{R})$ and its corresponding VA output $\tilde{\boldsymbol{v}}_\ell$. Both of them are likely to be large in the case when the candidate codeword is the transmitted one. Heuristically, we will set a threshold on $M_2(\hat{\boldsymbol{v}}^{(0)}_\ell)$ to check the correctness of the candidate codeword, as illustrated in \textbf{Example}~\ref{EX_THRESHOLD}.

\begin{figure}
  \centering
  \includegraphics[width=\figwidth]{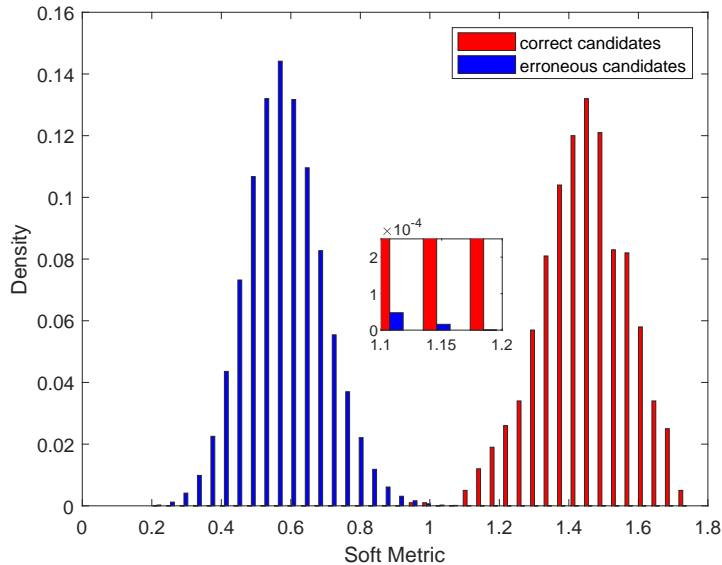}\\
  \caption{Statistical behavior of $M_2(\hat{\boldsymbol{v}}^{(0)}_\ell)$ in \textbf{Example}~\ref{EX_THRESHOLD}. The $16$-state $(2,1,4)$ TBCC defined by the polynomial generator matrix $G(D) = [27, 31]_8$ with information length $k=32$~($n=64$) is taken as the basic code. The setup of ${\rm SNR} = 3~{\rm dB}$ and $\ell_{\rm max}=64$ is considered.}\label{FIG_THRESHOLD}
\end{figure}

\begin{example}\label{EX_THRESHOLD}
The TBCC in \textbf{Example}~\ref{EX_LIST} is taken as the basic code. We set ${\rm SNR} = 3~{\rm dB}$ and $\ell_{\rm max}=64$. With the help of the histogram shown in Fig.~\ref{FIG_THRESHOLD}, we set a threshold $T$ to distinguish the correct candidate codeword from the erroneous one. The candidate codeword $\hat{\boldsymbol{v}}^{(0)}_\ell$ is treated to be correct only if $M_2(\hat{\boldsymbol{v}}^{(0)}_\ell) \geqslant T$, where $T$ is usually set large~(e.g., $T=1.2$ in this example) to reduce the probability that an erroneous candidate is mistaken as the correct one. The threshold $T$, depending on SNRs and coding parameters, can be determined off-line and stored for use in the decoding algorithm.
\end{example}

The list decoding algorithm, as summarized in Algorithm~\ref{DecodingAlgorithmW2}, is outlined as follows. The decoder employs the SLVA to compute the candidate codewords, which will be checked by~(\ref{SOFT_METRIC}) with a preset threshold, until finding a qualified one. If the list size reaches the maximum $\ell_{\rm max}$ and no candidate codeword is qualified, the decoder delivers $\hat{\boldsymbol{v}}^{(0)}_\ell$ with the maximum $M_2(\hat{\boldsymbol{v}}^{(0)}_\ell)$ as output.

%The successive cancellation decoding algorithm with decoding window $w=2$ for SRBO-CC is outlined as follows. For the first sub-frame, the decoder employs the SLVA to compute the decoding candidates, which will be checked by a statistical threshold, until finding a qualified one. If the list size reaches the maximum $\ell_{\rm max}$ and no decoding candidate is qualified, the decoder delivers $\hat{\boldsymbol{v}}^{(0)}_\ell$ with the maximum $M(\hat{\boldsymbol{v}}^{(0)}_\ell)$ as output. After removing the effect of the first sub-frame, the second sub-frame is then decoded in the same way. This process will be continued until all sub-frames are decoded. The detailed schedule for the decoding algorithm is summarized in Algorithm~\ref{DecodingAlgorithmW2}.

\begin{algorithm}\caption{Successive cancellation list decoding for the SRUMCC}\label{DecodingAlgorithmW2}
\begin{itemize}
  \item{\textbf{Global initialization:}} Set the threshold $T$. Assume that $\boldsymbol{y}^{(0)}$ has been received and set $\boldsymbol{z}^{(0)}=\boldsymbol{y}^{(0)}$.
  \item{\textbf{Sliding-window decoding:}}  For $0\leqslant t \leqslant L-1$, after receiving $\boldsymbol{y}^{(t+1)}$,
  \begin{enumerate}
    \item{\textbf{Local initialization:}} Set $M_{\rm max}=-\infty$ and $\ell = 1$.
    \item{\textbf{List:}}  While $M_{\rm max}\leqslant T$ and $\ell \leqslant \ell_{\rm max}$,
      \begin{enumerate}
        \item Perform the SLVA to find $\hat{\boldsymbol{v}}^{(t)}_\ell = {\rm SLVA}(\boldsymbol{z}^{(0)}, \ell)$ and compute $D(\hat{\boldsymbol{v}}^{(t)}_\ell,\boldsymbol{z}^{(0)})$.
        \item Flip the received vector $\boldsymbol{y}^{(t+1)}$, resulting in  $\boldsymbol{z}^{(1)} = \boldsymbol{y}^{(t+1)} \odot \phi(\hat{\boldsymbol{v}}^{(t)}_\ell\mathbf{R})$.
        \item Perform the VA to find $\tilde{\boldsymbol{v}}_\ell = {\rm VA}(\boldsymbol{z}^{(1)})$ and compute $D(\tilde{\boldsymbol{v}}_\ell,\boldsymbol{z}^{(1)})$.
        \item If $M_2(\hat{\boldsymbol{v}}^{(t)}_\ell)=D(\hat{\boldsymbol{v}}^{(t)}_\ell,\boldsymbol{z}^{(0)})+D(\tilde{\boldsymbol{v}}_\ell,\boldsymbol{z}^{(1)}) \geqslant M_{\rm max}$, replace $M_{\rm max}$ by $M_2(\hat{\boldsymbol{v}}^{(t)}_\ell)$ and $\hat{\boldsymbol{v}}^{(t)}_{\rm max}$ by $\hat{\boldsymbol{v}}^{(t)}_\ell$.
        \item Increment $\ell$ by one.
      \end{enumerate}

    \item{\textbf{Decision:}} Output $\hat{\boldsymbol{u}}^{(t)}$, the corresponding information vector to $\hat{\boldsymbol{v}}^{(t)}_{\rm max}$, as the decoding result of the $t$-th sub-frame.
    \item{\textbf{Cancellation:}} Remove the effect of the $t$-th sub-frame on the $(t+1)$-th sub-frame. That is, update $\boldsymbol{z}^{(0)}$ by computing
    \begin{displaymath}
    \boldsymbol{z}^{(0)} = \boldsymbol{y}^{(t+1)}\odot\phi(\hat{\boldsymbol{v}}^{(t)}_{\rm max}\mathbf{R}).
    \end{displaymath}
  \end{enumerate}
\end{itemize}
\end{algorithm}

\textbf{Remarks:}  The proposed list decoding algorithm is similar to the Feinstein's suboptimal decoder presented in~\cite[Theorem~18]{Polyanskiy2010Channel}, which is a conceptual algorithm to derive
 the performance bound for finite-length block codes.
 In Feinstein's decoder, all codewords are tested one-by-one in a preset order~(irrelevant to the received signal) by calculating the EDF.
 The first codeword with an EDF exceeding a fixed threshold is taken as the decoding output.
 This algorithm is rarely used in practice as we can imagine that the average number of tests to find the correct codeword is with the same order as the size of  the codebook.
In our algorithm, the codewords are tested serially in an order that is closely related to the received signal and determined by the SLVA on the first sub-frame.
The first codeword with an EDF exceeding an off-line learned tunable threshold is taken as the decoding output.
Evidenced by the simulation results, the transmitted codeword can be found with a small number of tests~(hence low complexity), especially in the high SNR region.
Compare to the fixed threshold, the tunable threshold is more attractive since the performance-complexity tradeoff can be achieved by adjusting the threshold.

It is worth pointing out that, besides the EDF, the likelihood function~(or the Euclidean distance under the assumption of AWGN channel) can also be employed as a metric for the test.
%In this paper, we define the soft metric based on the EDF for the convenience in threshold design.
In this paper, we define the soft metric based on the EDF since it can be applied to a general channel.
Another advantage of the EDF is the convenience in threshold design.
As we have discussed, the expectation of the EDF between the transmitted codeword and the received signal is the mutual information between the channel output and input.
Therefore, a rough threshold can be set directly based on the computable mutual information.
%\subsection{Improve Performance by Increasing Window}
%In the aforementioned algorithm, the decoding window is limited to $w=2$ to minimize the decoding latency, which is suitable for the system with strict latency constraint. If the constraint on the decoding latency is relaxed, however, we can increase the decoding window to improve the performance. In the following, we discuss how to estimate $\boldsymbol{v}^{(0)}$ from $\boldsymbol{y}^{(0)}$, $\boldsymbol{y}^{(1)}$ and $\boldsymbol{y}^{(2)}$.

\section{Performance and Complexity Analysis}\label{SEC_4}
\subsection{Upper Bound}
In this subsection, we derive an upper bound on ${\rm fER}_0$ under the ML decoding. Because of the linearity of the code, we assume that all zero codeword is transmitted. The ML decoder selects $\hat{\boldsymbol{v}}^{(0)}$ as output such that the codewords $(\hat{\boldsymbol{v}}^{(0)},\hat{\boldsymbol{v}}^{(1)})$ maximize $P(\boldsymbol{y}^{(0)}\boldsymbol{y}^{(1)}|\boldsymbol{v}^{(0)}\boldsymbol{v}^{(1)})$. The ML decoding is successful if $\hat{\boldsymbol{v}}^{(0)} = \boldsymbol{0}$ and an error occurs if $\hat{\boldsymbol{v}}^{(0)}\neq \boldsymbol{0}$. Note that $\hat{\boldsymbol{v}}^{(0)}$ can be correct even if $\hat{\boldsymbol{v}}^{(1)}\neq \boldsymbol{0}$. The ${\rm fER}_0$ can be upper bounded by
\begin{align}
{\rm fER}_0 &= {\rm Pr}\left\{
\bigcup_{\begin{subarray}{c}\boldsymbol{v}^{(0)}\neq \boldsymbol{0}\\ \boldsymbol{v}^{(1)}\end{subarray}}
(\boldsymbol{v}^{(0)},\boldsymbol{v}^{(1)}){\rm ~is~most~likely}
\right\} \nonumber\\
&\leqslant {\rm Pr}\left\{
\bigcup_{\begin{subarray}{c}\boldsymbol{v}^{(0)}\neq \boldsymbol{0}\\ \boldsymbol{v}^{(1)}\end{subarray}}
(\boldsymbol{v}^{(0)},\boldsymbol{v}^{(1)}){\rm ~is~more~likely~than~}(\boldsymbol{0},\boldsymbol{0})
\right\}\nonumber\\
&\leqslant \sum_{\begin{subarray}{c}\boldsymbol{v}^{(0)}\neq \boldsymbol{0}\\ \boldsymbol{v}^{(1)}\end{subarray}}
{\rm Pr}\left\{
(\boldsymbol{v}^{(0)},\boldsymbol{v}^{(1)}){\rm ~is~more~likely~than~}(\boldsymbol{0},\boldsymbol{0})
\right\}.
\end{align}

%This bound is similar to the well-known union bound and closely related to the weight distribution of the truncated code
This bound is indeed the well-known union bound and can be calculated by deriving the weight distribution of the truncated code
\begin{equation}
\mathscr{C}^{(0,1)} = \left\{(\boldsymbol{c}^{(0)}, \boldsymbol{c}^{(1)}) \bigg|
\begin{array}{c}
\boldsymbol{c}=(\boldsymbol{c}^{(0)}, \cdots, \boldsymbol{c}^{(L)}) {\rm~is~a~coded}\\{\rm~sequence~with~} \boldsymbol{c}^{(0)} \neq \boldsymbol{0}
\end{array}
\right\}.
\end{equation}
Let $A(X)$ be the WEF of the basic code $\mathscr{C}\backslash{\boldsymbol{0}}$~(all non-zero codewords). Then the {\em ensemble} WEF of the truncated code $\mathscr{C}^{(0,1)}$ with $\mathbf{R}$ being totally random is given by
\begin{equation}
B(X) = 2^{-n+k}(1+X)^n A(X)=\sum_{w=1}^{2n}B_wX^w.
\end{equation}
The upper bound on ${\rm fER}_0$ under the ML decoding is given by
\begin{equation}\label{bound_subfer}
{\rm fER}_0 \leqslant \sum_{w=1}^{2n}B_wQ\left(\sqrt{\frac{w}{\sigma^2}}\right),
\end{equation}
where $\sigma^2$ is the variance of the noise. Note that the bounding technique in~\cite{New2013Ma}, which is based on triplet-wise error probabilities, can also be applied here to tighten the upper bound in low SNR region. In this paper, we simply employ the union bound since we are interested in the performance in high SNR region, which can be well predicted by the union bound.

\subsection{Lower Bound and Extended Windowed Decoding}\label{SUBSEC_DECW3}
Obviously, in the list decoding, the first sub-frame can be decoded correctly only if the transmitted codeword is included in the list. Therefore, the ${\rm fER}_0$ performance is not better than the list decoding performance of the basic code, which can be regarded as a lower bound and obtained by simulating the list decoding of the basic code. \textbf{Example}~\ref{EX_LOWERBOUND} is presented to illustrate the lower bound on ${\rm fER}_0$.

\begin{figure}
  \centering
  \includegraphics[width=\figwidth]{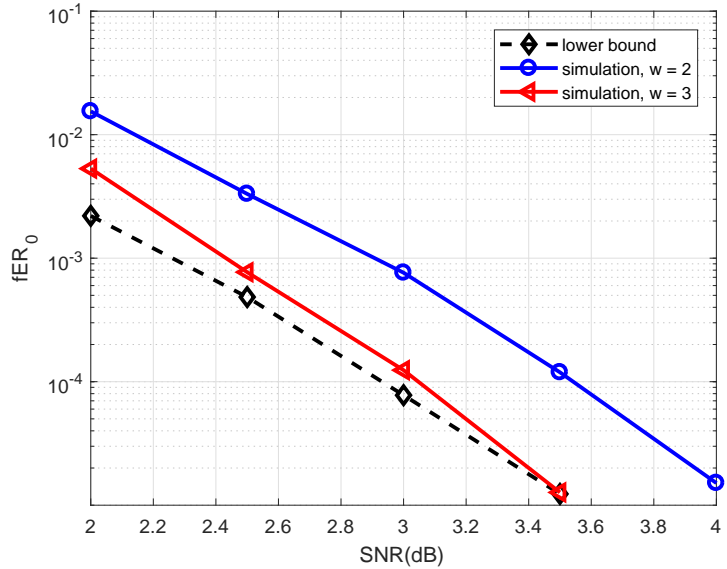}\\
  \caption{Performance of the list decoding with $w=2$ and $w=3$ in \textbf{Example}~\ref{EX_LOWERBOUND}. The basic code is the $16$-state $(2,1,4)$ convolutional code with information length $k=32$~($n=64$), which is truncated without termination and defined by the polynomial generator matrix $G(D) = [27, 31]_8$. The list size is $\ell_{\rm max} = 64$ and the thresholds are set properly based on the statistical behavior of the EDF. The corresponding lower bound is also plotted.}\label{FIG_LOWERBOUND}
\end{figure}

%The ${\rm fER}_0$ performance of the statistical learning aided decoding with $w=2$ and $w=3$~(discussed in the following) are shown in Fig.~\ref{FIG_LOWERBOUND}. The corresponding lower bound is also plotted.
\begin{example}\label{EX_LOWERBOUND}
The basic code is the $16$-state $(2,1,4)$ convolutional code with information length $k=32$~($n=64$), which is truncated without termination and defined by the polynomial generator matrix $G(D) = [27, 31]_8$. The list size is $\ell_{\rm max} = 64$ and the thresholds are set properly based on the statistical behavior of the EDF. The ${\rm fER}_0$ performance of the list decoding are shown in Fig.~\ref{FIG_LOWERBOUND}, where ``$w = 2$'' corresponds to Algorithm~\ref{DecodingAlgorithmW2}. The corresponding lower bound is also plotted. We see that Algorithm~\ref{DecodingAlgorithmW2} performs about $0.5~{\rm dB}$ away from the lower bound, implying that the statistical check is not always able to identify the transmitted codeword in the list. This gap can be narrowed, however, if the constraint on complexity and latency is relaxed. Indeed, we can extend Algorithm~\ref{DecodingAlgorithmW2}, which recovers $\boldsymbol{v}^{(0)}$ from $\boldsymbol{y}^{(0)}$ and $\boldsymbol{y}^{(1)}$, to improve the performance by recovering $\boldsymbol{v}^{(0)}$ from $\boldsymbol{y}^{(0)}$, $\boldsymbol{y}^{(1)}$ and $\boldsymbol{y}^{(2)}$, as shown by the curve ``$w = 3$'' in Fig.~\ref{FIG_LOWERBOUND}. The details of such an extension is omitted here, while the basic idea is described below.
\end{example}

%From Fig.~\ref{FIG_LOWERBOUND}, we see that the aforementioned statistical learning aided decoding algorithm with $w=2$ performs about $0.5~{\rm dB}$ away from the lower bound, implying that the decoder is not able to identify the transmitted codeword even the transmitted codeword is included in the list. If the constraint on complexity and latency is relaxed, however, we can extend the decoding window to improve the performance. In the following, we discuss how to estimate $\boldsymbol{v}^{(0)}$ from $\boldsymbol{y}^{(0)}$, $\boldsymbol{y}^{(1)}$ and $\boldsymbol{y}^{(2)}$.

After receiving $\boldsymbol{y}^{(0)}$ and $\boldsymbol{y}^{(1)}$, the decoder first attempts to recover $\boldsymbol{v}^{(0)}$ by Algorithm~\ref{DecodingAlgorithmW2}. In the case when the decision on $\boldsymbol{v}^{(0)}$ is not that confident, we keep a list of candidates for further processing. For each candidate $\hat{\boldsymbol{v}}^{(0)}$, we perform Algorithm~\ref{DecodingAlgorithmW2} to find $\hat{\boldsymbol{v}}^{(1)}$ and $\hat{\boldsymbol{v}}^{(2)}$ from $\boldsymbol{y}^{(1)}$ and $\boldsymbol{y}^{(2)}$. Finally, we select $\hat{\boldsymbol{v}}^{(0)}$ such that $(\hat{\boldsymbol{v}}^{(0)},\hat{\boldsymbol{v}}^{(1)},\hat{\boldsymbol{v}}^{(2)})$ is the most likely candidate with respect to $(\boldsymbol{y}^{(0)},\boldsymbol{y}^{(1)},\boldsymbol{y}^{(2)})$.

\subsection{Decoding Complexity}
%Consider a basic code $\mathscr{C}[n,k]$ with a trellis representation of $s$ states.
In this subsection, taking the add-compare-select operation~(the basic operation in both the VA and the SLVA) as an atomic operation, we analyze the complexity of the list decoding with empirical divergence test. Assume that the basic code $\mathscr{C}[n,k]$ has a trellis representation with $s$ states. To find the best candidate codeword by the SLVA~(equivalently, by the VA), $sn$ operations are needed. With the $(\ell-1)$-th best candidate codeword known, only $n$ operations are needed to find the $\ell$-th best candidate codeword by the SLVA.

Let $\bar{\ell}\leqslant \ell_{\rm max}$ be the average list size. Then the SLVA requires on average $sn+(\bar{\ell}-1)n$ operations. For each candidate, the VA is employed to calculate the soft metric, which needs $\bar{\ell} sn$ operations. Hence the total operations for decoding each sub-frame is given by
\begin{equation}
\#\text{Operations}=(s+\bar{\ell}-1+\bar{\ell} s)n.
\end{equation}

We see that the complexity is dominated by $\bar{\ell} sn$. For fixed $n$, to reduce the complexity, we can reduce the average list size $\bar{\ell}$ by tuning down the threshold. %Another way to reduce the number of states $s$ by employing the sub-optimal decoding algorithm, such as the WAVA for tail-biting convolutional codes.

\section{Simulation Results}\label{SEC_5}
In this section, all simulations are conducted by assuming BPSK modulation and AWGN {channels}. The SRUMCCs are terminated every $L=49$ blocks. All codes are decoded by Algorithm~\ref{DecodingAlgorithmW2} with the maximum list size $\ell_{\rm max}=64$ and properly thresholds obtained based on the statistical behavior of the EDF, unless otherwise specified. All upper bounds taken as the benchmarks are derived by combining~(\ref{bound_fer}) and~(\ref{bound_subfer}).

\subsection{Impact of Sub-frame Length on the Performance}
\begin{figure}
  \centering
  \includegraphics[width=\figwidth]{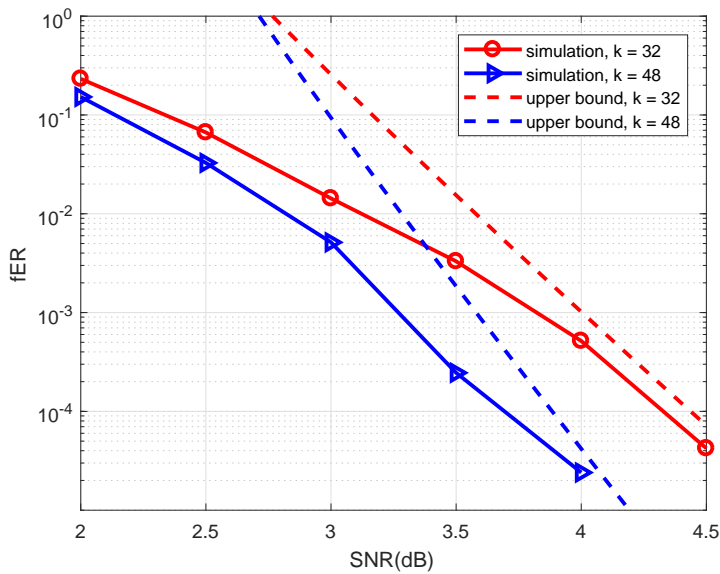}\\
  \caption{Performance of the SRUMCCs in \textbf{Example}~\ref{EXEX_PARAM}. The basic code is the $16$-state $(2,1,4)$ convolutional code, which is truncated without termination and defined by the polynomial generator matrix $G(D) = [27, 31]_8$. Different sub-frame information lengths $k=32,48$ and different maximum list sizes $\ell_{\rm max}=64,128$ are considered.}\label{FIG_EXEX_PARAM}
\end{figure}

%It should be pointed out that
\begin{example}\label{EXEX_PARAM}
The basic code is the $16$-state $(2,1,4)$ convolutional code, which is truncated without termination and defined by the polynomial generator matrix $G(D) = [27, 31]_8$. Different sub-frame information lengths $k=32,48$ and different maximum list sizes $\ell_{\rm max}=64,128$ are considered. The $\rm fER$ is shown in Fig.~\ref{FIG_EXEX_PARAM}. The upper bounds indicate that the ML performance of the SRUMCCs can be improved by increasing $k$~(hence the decoding delay). It is also worth pointing out that a larger $k$ usually requires a larger maximum list size $\ell_{\rm max}$. For $k=32$, the performance curve with $\ell_{\rm max}=64$ matches that with $\ell_{\rm max}=128$, indicating that the performance is saturated with $\ell_{\rm max}=64$. However, for $k=48$, the performance can be improved by increasing the maximum list size from $\ell_{\rm max}=64$ to $\ell_{\rm max}=128$.
%We see that the performance can be improved by increasing $k$~(hence the decoding delay). It is worth pointing out that a larger $k$ usually requires a larger maximum list size $\ell_{\rm max}$.
% with large $k$, a large maximum list size $\ell_{\rm max}$ is needed.
\end{example}

\subsection{Tradeoff Between Performance and Complexity}
\begin{figure}
  \centering
  \includegraphics[width=\figwidth]{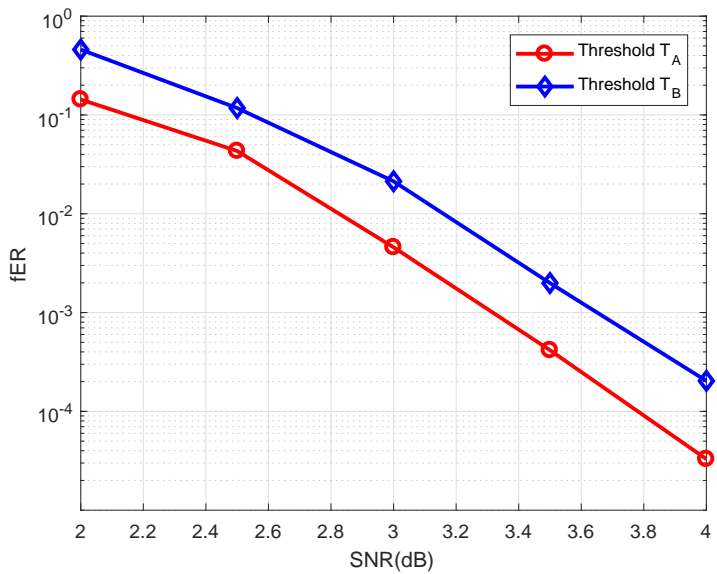}\\
  \caption{Performance of the SRUMCC in \textbf{Example}~\ref{EXEX_THRES}. The $16$-state $(2,1,4)$ TBCC defined by the polynomial generator matrix $G(D) = [27, 31]_8$ is taken as the basic code. The sub-frame information length is $k=32$. We consider two sets of thresholds $T_A$ and $T_B$ specified in Table~\ref{TAB_EXEX_THRES}.}\label{FIG_EXEX_THRES}
\end{figure}

\begin{example}\label{EXEX_THRES}
%Consider the code in \textbf{Example}~\ref{EX_3} again. Another set of thresholds $T_B = 0.95,1.0,1.05,1.1,1.15$ are chosen for ${\rm SNR} = 2.0,2.5,3.0,3.5,4.0$, respectively. The fER is shown in Fig.~\ref{FIG_SIM2}, while the average list size needed for decoding a sub-frame is shown in Table~\ref{TAB_SIM2}. It can be seen that the complexity~(average list size), at the cost of performance loss, can be reduced by tuning down the threshold. For example, at ${\rm SNR} = 4~{\rm dB}$, the computational complexity~(average list size) can be reduced more than $10$ times if a performance degradation~(fER deterioration) is tolerated from $10^{-5}$ to $10^{-4}$.
The $16$-state $(2,1,4)$ TBCC defined by the polynomial generator matrix $G(D) = [27, 31]_8$ is taken as the basic code. The sub-frame information length is $k=32$. We consider two sets of thresholds $T_A$ and $T_B$ specified in Table~\ref{TAB_EXEX_THRES}. The $\rm fER$ is shown in Fig.~\ref{FIG_EXEX_THRES}, while the average list sizes needed for decoding a sub-frame are shown in Table~\ref{TAB_EXEX_THRES}. We see that the complexity~(average list size), at the cost of performance loss, can be reduced by tuning down the threshold. For example, at ${\rm SNR} = 4~{\rm dB}$, the computational complexity~(average list size) can be reduced more than $10$ times if a performance degradation~($\rm fER$ deterioration) is tolerated from $10^{-5}$ to $10^{-4}$.
\end{example}

\begin{table}[tp]
  \centering
  \caption{Average list sizes needed for $T_A$ and $T_B$}
  \begin{tabular}{|l|l|l|l|l|l|l|}\hline
  ${\rm SNR}$             & 2.0 & 2.5 & 3.0 & 3.5 & 4.0\\\hline
  $T_A$                   & 1.3 &1.35 & 1.4 &1.45 & 1.5\\\hline
  $T_B$                   &0.95 & 1.0 &1.05 & 1.1 &1.15\\\hline
  list size for $T_A$     & 38  & 30  & 23  & 18  & 14 \\\hline
  list size for $T_B$     & 25  & 8.2 & 2.6 & 1.3 & 1.1\\\hline
  \end{tabular}\label{TAB_EXEX_THRES}
\end{table}

\subsection{Performance with Different Rates}
\begin{figure}
  \centering
  \includegraphics[width=\figwidth]{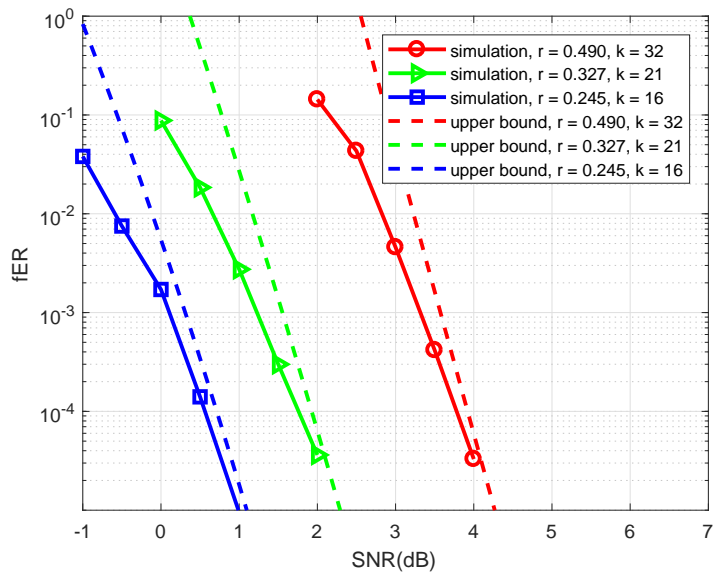}\\
  \caption{Performance of the SRUMCCs in \textbf{Example}~\ref{EXEX_RATE}. The $16$-state $(2,1,4)$ TBCC defined by the polynomial generator matrix $G(D) = [27, 31]_8$, the $(3,1,4)$ TBCC defined by the polynomial generator matrix $G(D) = [25, 33, 37]_8$ and the $(4,1,4)$ TBCC defined by the polynomial generator matrix $G(D) = [25, 27, 33, 37]_8$ are taken as the basic code. The sub-frame information lengths and the total rates are specified in the legends.}\label{FIG_EXEX_RATE}
\end{figure}

\begin{example}\label{EXEX_RATE}
The $16$-state $(2,1,4)$ TBCC defined by the polynomial generator matrix $G(D) = [27, 31]_8$, the $(3,1,4)$ TBCC defined by the polynomial generator matrix $G(D) = [25, 33, 37]_8$ and the $(4,1,4)$ TBCC defined by the polynomial generator matrix $G(D) = [25, 27, 33, 37]_8$ are taken as the basic code. The sub-frame information lengths and the total rates are specified in the legends. The $\rm fER$ is shown in Fig.~\ref{FIG_EXEX_RATE}. We see that the SRUMCCs can support a wide range of code rates by simply choosing the basic code with the desired rate. %adapting the basic code to the desired rate.
\end{example}

\subsection{Comparison with Sequential Decoding}
\begin{figure}
  \centering
  \includegraphics[width=\figwidth]{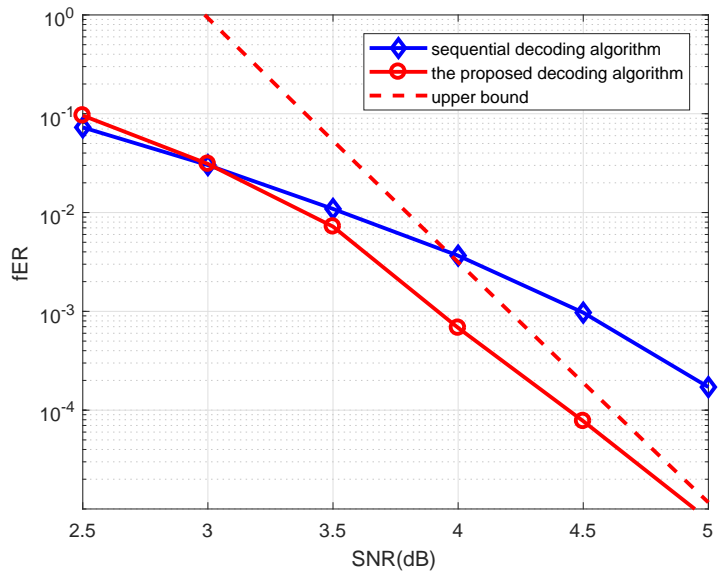}\\
  \caption{Performance of the SRUMCC in \textbf{Example}~\ref{EXEX_DECODING}. The Cartesian product of Reed-Muller code ${\rm RM}[8,4]^8$ is taken as the basic code. The sub-frame information length is $k=32$. For comparison, the same code is also decoded by the sequential decoding~\cite{Ma2018SRBOCC} with the same decoding window and a stack of size 20000.}\label{FIG_EXEX_DECODING}
\end{figure}

\begin{example}\label{EXEX_DECODING}
%We set $k=32$ and $\ell_{\rm max}=64$. A set of thresholds $T_A = 1.3,1.35,1.4,1.45,1.5$ are chosen for ${\rm SNR} = 2.0,2.5,3.0,3.5,4.0$, respectively. The fER is shown in Fig.~\ref{FIG_SIM1}. For comparison, we have also redrawn the performance curves of the polar code~\cite{Wu2016OSDPolar} with length 128~(the same decoding delay as the SRBO-CC). We observe that the SRBO-CC with successive cancellation decoding is competitive with the polar code.
The Cartesian product of Reed-Muller code ${\rm RM}[8,4]^8$ is taken as the basic code. The sub-frame information length is $k=32$. For comparison, the same code is also decoded by the sequential decoding~\cite{Ma2018SRBOCC} with the same decoding window and a stack of size 20000. The $\rm fER$ is shown in Fig.~\ref{FIG_EXEX_DECODING}. We see that the proposed list decoding algorithm outperforms the sequential decoding algorithm in high SNR region.
\end{example}

\subsection{Comparison with Other Codes}
\begin{figure}
  \centering
  \includegraphics[width=\figwidth]{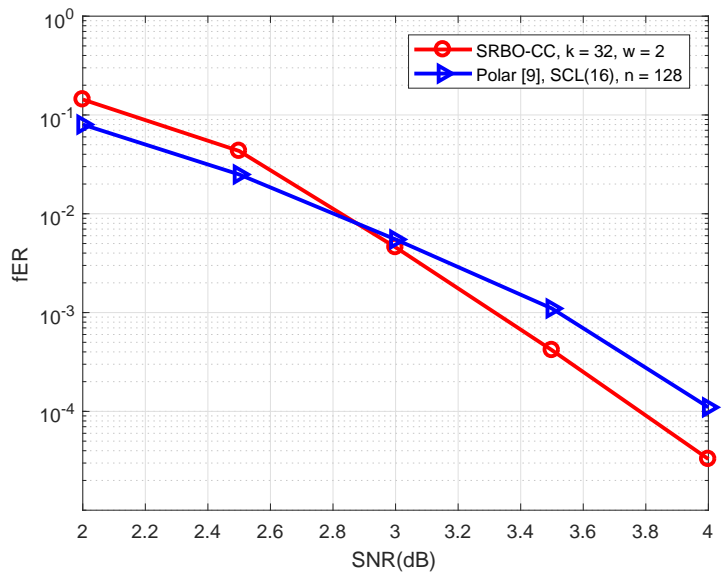}\\
  \caption{Performance of the SRUMCC in \textbf{Example}~\ref{EXEX_COMP}. The $16$-state $(2,1,4)$ TBCC defined by the polynomial generator matrix $G(D) = [27, 31]_8$ is taken as the basic code. The sub-frame information length is $k=32$. For comparison, we have also redrawn the performance curve of the polar code~\cite{Hashemi2018RMP} without CRC. The coding length of the polar code is 128~(the same decoding delay as the SRUMCC). Note that ``SCL(16)'' represents the successive cancellation list algorithm~\cite{Tal2011SCL} with list size 16.}\label{FIG_EXEX_COMP}
\end{figure}

\begin{example}\label{EXEX_COMP}
%We set $k=32$ and $\ell_{\rm max}=64$. A set of thresholds $T_A = 1.3,1.35,1.4,1.45,1.5$ are chosen for ${\rm SNR} = 2.0,2.5,3.0,3.5,4.0$, respectively. The fER is shown in Fig.~\ref{FIG_SIM1}. For comparison, we have also redrawn the performance curves of the polar code~\cite{Wu2016OSDPolar} with length 128~(the same decoding delay as the SRBO-CC). We observe that the SRBO-CC with successive cancellation decoding is competitive with the polar code.
The $16$-state $(2,1,4)$ TBCC defined by the polynomial generator matrix $G(D) = [27, 31]_8$ is taken as the basic code. The sub-frame information length is $k=32$. For comparison, we have also redrawn the performance curve of the polar code~\cite{Hashemi2018RMP} without CRC. The coding length of the polar code is 128~(the same decoding delay as the SRUMCC). The $\rm fER$ is shown in Fig.~\ref{FIG_EXEX_COMP}, where ``SCL(16)'' represents the successive cancellation list algorithm~\cite{Tal2011SCL} with list size 16. We see that the SRUMCC with  list decoding is competitive with the polar code.
\end{example}

\section{Conclusion}\label{SEC_6}
In this paper, we have presented more details on the SRUMCCs, which can be decoded by successive cancellation list decoding with empirical divergence test. The decoder outputs serially a list of decoding candidates and identifies the correct one by a statistical threshold, which can be designed based on the statistical behavior of the EDF. The performance-complexity tradeoff and the performance-delay tradeoff can be achieved by adjusting the statistical threshold and the decoding window size. A closed-form upper bound based on the weight enumerating function was derived to analyze the performance in high SNR region. Simulation results showed that the proposed list decoding outperforms the sequential decoding in high SNR region and that under the constraint of equivalent decoding delay, the SRUMCCs have comparable performance with the polar codes.

%\section*{Acknowledgment}

\ifCLASSOPTIONcaptionsoff
  \newpage
\fi

\bibliographystyle{IEEEtran}
\bibliography{IEEEabrv,bibfile}

\end{document}